\input amstex
\documentstyle{amsppt}
%
%
\nopagenumbers
\def\negskp{\hskip -2pt}
\def\compos{\,\raise 1pt\hbox{$\sssize\circ$} \,}
\def\divr{\operatorname{div}}
\def\tr{\operatorname{tr}}
\def\const{\operatorname{const}}
\accentedsymbol\ty{\tilde y}
\pagewidth{360pt}
\pageheight{606pt}
\leftheadtext{S.~F.~Lyuksyutov, R.~A.~Sharipov}
\rightheadtext{Note on kinematics, dynamics,
and thermodynamics \dots}
\topmatter
\title
Note on kinematics, dynamics,
and thermodynamics of plastic glassy media.
\endtitle
\author
S.~F.~Lyuksyutov, R.~A.~Sharipov
\endauthor
\address Physics Department, Ayer Hall, 213,
University of Akron, Akron, OH 44325
\endaddress
\email sfl\@physics.uakron.edu
\endemail
\address Mathematics Department, Bashkir State University,
Frunze street 32, 450074 Ufa, Russia
\endaddress
\email \vtop to 20pt{\hsize=280pt\noindent
R\_\hskip 1pt Sharipov\@ic.bashedu.ru\newline
r-sharipov\@mail.ru\vss}
\endemail
\urladdr
http:/\negskp/www.geocities.com/r-sharipov
\endurladdr
\abstract
    Unified geometric approach to describing kinematics of
elastic and plastic deformations of continuous media is
suggested. On the base of this approach we study mechanical
deformations, viscous flow, and heat transport in glassy
plastic media. As a result we derive appropriate differential
equations for these phenomena in a form applicable to liquids,
elastic solids, and to plastic solid materials as well.
\endabstract
\endtopmatter
\loadbold
\TagsOnRight
\document
\head
1. Introduction.
\endhead
    AFM-assisted\footnotemark\ electrostatic nanolithography
(AFMEN) was suggested in \cite{1} and \cite{2}. It is new
technique for nano-scale patterns formation on planar polymer
films. \footnotetext{Here AFM is abbreviation for Atomic Force
Microscope.}By this paper we start series of papers aimed to
prepare theoretical background for further numeric simulation
of AFMEN process. First of all we consider kinematics (i\.\,e\.
time-dependent geometry) of deformations and apply differential
geometry to study it. Our main result is a method for separating
elastic and plastic deformations within general nonlinear
deformation tensor. Then we build this method into the standard
framework of balance equations traditionally used to describe
dynamics and thermodynamics of moving continuous media. Applying
electrostatic field, surface phenomena, and phase transitions
will be considered in separate papers.\par
    As far as our technique is concerned, we use curvilinear
coordinates from the very beginning. This looks a little bit
tricky, but in this way we make transparent tensorial nature of
all quantities we use. Moreover, the problem of AFMEN process
simulation, which we are going to study numerically, has obvious
cylindrical symmetry. So we prepare use of cylindrical coordinates
in our future calculations. For reader's convenience in section~2
just below we resume some well-known facts concerning curvilinear
coordinates.\par
\head
2. Moving frame of curvilinear coordinates.
\endhead
     Let $x^1,\,x^2,\,x^3$ be three Cartesian coordinates of a point.
Below we use both upper and lower indices, following traditions of
tensorial analysis (see \cite{3}). Curvilinear coordinates $y^1,\,y^2,
\,y^3$ of a point are usually introduced by functions
$$
\hskip -2em
\cases x^1=x^1(y^1,y^2,y^3),\\
x^2=x^2(y^1,y^2,y^3),\\
x^3=x^3(y^1,y^2,y^3).
\endcases
\tag2.1
$$
Functions \thetag{2.1} define transition to curvilinear coordinates.
Transition back to Cartesian coordinates is determined by similar
three functions
$$
\hskip -2em
\cases x^1=x^1(y^1,y^2,y^3),\\
x^2=x^2(y^1,y^2,y^3),\\
x^3=x^3(y^1,y^2,y^3).
\endcases
\tag2.2
$$
Maps \thetag{2.1} and \thetag{2.2} are inverse to each other. Therefore
if we define their Jacobi matrices $S$ and $T$ by partial derivatives
$$
\xalignat 2
&\hskip -2em
S^{\,i}_j=\frac{\partial x^i}{\partial y^j},
&&T^{\,i}_j=\frac{\partial y^i}{\partial x^j},
\tag2.3
\endxalignat
$$
they are also inverse to each other: $T=S^{-1}$. Matrix $S$ is called
{\it direct transition matrix}, while $T$ is called {\it inverse transition
matrix}.\par
    Let $\bold e_1,\,\bold e_2,\,\bold e_3$ be constant frame of Cartesian
coordinates $x^1,\,x^2,\,x^3$, i\.\,e\. these are three base vectors directed
along three coordinate axes. Then, using functions \thetag{2.1}, we can
define vectorial function $\bold r=\bold r(y^1,y^2,y^3)$:
$$
\hskip -2em
\bold r=\sum^3_{i=1}x^i(y^1,y^2,y^3)\,\bold e_i.
\tag2.4
$$
This is radius-vector of a point expressed through curvilinear coordinates.
Differentiating \thetag{2.4} with respect to $y^1,\,y^2,\,y^3$, we get
three vector-functions:
$$
\hskip -2em
\bold E_i=\bold E_i(y^1,y^2,y^3)=\frac{\partial\bold r}{\partial y^i},
\qquad i=1,2,3.
\tag2.5
$$
Vectors $\bold E_1,\,\bold E_2,\,\bold E_3$ form a frame of curvilinear
coordinates $y^1,\,y^2,\,y^3$. This is moving frame, since vectors
$\bold E_1,\,\bold E_2,\,\bold E_3$ depend on coordinates of a point
to which they are attached. Transition matrices \thetag{2.3} relate
moving frame $\bold E_1,\,\bold E_2,\,\bold E_3$ to constant frame
$\bold e_1,\,\bold e_2,\,\bold e_3$ of Cartesian coordinates and vice
versa:
$$
\xalignat 2
&\bold E_i=\sum^3_{q=1}S^{\,q}_i\,\bold e_q,
&&\bold e_i=\sum^3_{q=1}T^{\,q}_i\,\bold E_q.
\endxalignat
$$\par 
    Mutual scalar products of frame vectors \thetag{2.5} define
fundamental tensor of our space. This is metric tensor $\bold g$
with components:
$$
\hskip -2em
g_{ij}=(\bold E_i,\,\bold E_j).
\tag2.6
$$
Matrix \thetag{2.6} also is known as {\it Gram matrix} of moving frame
$\bold E_1,\,\bold E_2,\,\bold E_3$. Inverse matrix to \thetag{2.6}
define another fundamental tensor. This is {\it dual metric tensor};
by tradition it is denoted by the same letter $\bold g$, but for its
components upper indices are used: $g^{ij}$. Being mutually inverse,
metric tensors are related by formula
$$
\pagebreak
\sum^3_{q=1}g^{iq}\,g_{qj}=\delta^i_j.
$$
Here $\delta^i_j$ are Kronecker symbols. They determine components of
unit matrix:
$$
\delta^i_j=\cases 1\text{\ \  for \ }i=j,\\
0\text{\ \  for \ }i\neq j.\endcases
$$
\head
3. Tensor fields in curvilinear coordinates.
\endhead
    All tensor fields in curvilinear coordinates $y^1,\,y^2,\,y^3$
are referenced to moving frame $\bold E_1,\,\bold E_2,\,\bold E_3$.
This determine some features of their differentiation. Thus, if $\varphi=
\varphi(y^1,y^2,y^3)$ is a scalar field, then applying gradient operator
$\nabla$ to it we get three components defined by partial derivatives:
$$
\hskip -2em
\nabla_i\varphi=\frac{\partial\varphi}{\partial y^i},\qquad
i=1,2,3.
\tag3.1
$$
However, unlike \thetag{3.1}, applying gradient operator to tensor
field $\bold X$ is more complicated procedure. For components of
resulting tensor field $\nabla\bold X$ we have formula
$$
\hskip -2em
\gathered
\nabla_{\!m}X^{i_1\ldots\,i_r}_{j_1\ldots\,j_s}=\frac{\partial
X^{i_1\ldots\,i_r}_{j_1\ldots\,j_s}}{\partial y^m}\,+\\
+\sum^r_{k=1}\sum^3_{a_k=1}\!\Gamma^{i_k}_{m\,a_k}\,X^{i_1\ldots\,
a_k\ldots\,i_r}_{j_1\ldots\,\ldots\,\ldots\,j_s}
-\sum^s_{k=1}\sum^3_{b_k=1}\!\Gamma^{b_k}_{m\,j_k}
X^{i_1\ldots\,\ldots\,\ldots\,i_r}_{j_1\ldots\,b_k\ldots\,j_s}.
\endgathered
\tag3.2
$$
Here $\Gamma^k_{ij}$ is three-dimensional array of {\it connection
components} or {\it Christoffel symbols}. They are defined by metric
tensor $\bold g$ according to the following formula:
$$
\hskip -2em
\Gamma^k_{ij}=\sum^3_{s=1}\frac{g^{ks}}{2}\left(\frac{\partial
g_{is}}{\partial y^j}+\frac{\partial g_{sj}}{\partial y^i}-
\frac{\partial g_{ij}}{\partial y^s}\right).
\tag3.3
$$
Formulas \thetag{3.2} and \thetag{3.3} are well-known in differential
geometry (see \cite{3}).\par
    In addition to metric tensor \thetag{2.6} and connection components
\thetag{3.3} there are also {\it volume tensor} $\boldsymbol\omega$ and
{\it dual volume tensor} denoted by the same symbol:
$$
\xalignat 2
&\hskip -2em
\omega_{ijk}=\sqrt{\det\bold g}\ \varepsilon_{ijk},
&&\omega^{ijk}=\frac{\varepsilon^{ijk}}{\sqrt{\det\bold g}},
\tag3.4
\endxalignat
$$
Here $\varepsilon_{ijk}=\varepsilon^{ijk}$ are Levi-Civita symbols.
They are defined as follows:
$$
\hskip -2em
\varepsilon_{ijk}=\varepsilon^{ijk}=\cases
\ 0 &\text{if $i=j$, $j=k$, or $k=i$;}\\
\ 1 &\text{if $(ijk)$ is even transposition of $(123)$;}\\
-1 &\text{if $(ijk)$ is odd transposition of $(123)$.} 
\endcases
\tag3.5
$$
Levi-Civita symbols \thetag{3.5} do not form a tensor. However,
supplying scalar factors to them, we get two tensor fields
\thetag{3.4}.\par
\head
4. Deformation of continuous medium.
\endhead
    Suppose that our space is filled with medium of some kind.
Deformation of medium is due to the displacement of its points.
\pagebreak Suppose that the point with coordinates $\ty^1,\,\ty^2,\,
\ty^3$ has moved to the point with coordinates $y^1,\,y^2,\,y^3$.
This situation is expressed by the following three functions:
$$
\hskip -2em
\cases y^1=y^1(t,\ty^1,\ty^2,\ty^3),\\
y^2=y^2(t,\ty^1,\ty^2,\ty^3),\\
y^3=y^3(t,\ty^1,\ty^2,\ty^3).
\endcases
\tag4.1
$$
Argument $t$ in \thetag{4.1} is responsible for time evolution
of displacement. Time derivatives of these functions determine
velocity vector components:
$$
\hskip -2em
v^i=\dot y^i=\frac{\partial y^i}{\partial t},\qquad
i=1,2,3.
\tag4.2
$$
Velocity vector $\bold v$ itself is calculated as a sum representing
its expansion in moving frame at the point with coordinates $y^1,\,y^2,
\,y^3$:
$$
\hskip -2em
\bold v=\sum^3_{i=1}v^i\,\bold E_i.
\tag4.3
$$
Functions \thetag{4.1} define time-dependent map from space to space.
Let's denote it by $\tau$. Then inverse map $\tau^{-1}$ is given by
similar functions
$$
\hskip -2em
\cases \ty^1=\ty^1(t,y^1,y^2,y^3),\\
\ty^2=\ty^2(t,y^1,y^2,y^3),\\
\ty^3=\ty^3(t,y^1,y^2,y^3).
\endcases
\tag4.4
$$
Quantities $v^i$, as defined in \thetag{4.2}, are functions of
coordinates $\ty^1,\,\ty^2,\,\ty^3$ marking initial position of a point
of medium. Using \thetag{4.4} we can convert them to the functions of
coordinates $y^1,\,y^2,\,y^3$ marking current actual position of that
point:
$$
\hskip -2em
v^i=v^i(t,y^1,y^2,y^3)=\frac{\partial y^i}{\partial t}\compos\tau^{-1}:
\tag4.5
$$
Time dependent maps \thetag{4.1} and \thetag{4.4} define two Jacobi
matrices $\tilde S$ and $\tilde T$:
$$
\xalignat 2
&\hskip -2em
\tilde S^{\,i}_j=\frac{\partial y^i}{\partial\ty^j},
&&\tilde T^{\,i}_j=\frac{\partial\ty^i}{\partial y^j}.
\tag4.6
\endxalignat
$$
Matrices \thetag{4.6} are inverse to each other and quite similar to
that of \thetag{2.3}. However, in contrast to matrices $S$ and $T$,
these two matrices depend on $t$ and describe physical state of our
continuous medium (more precisely, they describe the state of
deformation). Matrix $\tilde T$ has proper arguments $t,\,y^1,\,y^2,
\,y^3$, while arguments of $\tilde S$ should be corrected by inverse
map \thetag{4.4}, i\.\,e\. $\tilde S\to\tilde S\compos\tau^{-1}$.\par
    Let's take vectors of moving frame $\bold E_1,\,\bold E_2,\,
\bold E_3$ at the point with coordinates $y^1,\,y^2,\,y^3$ and send
them to the point with coordinates $\ty^1,\,\ty^2,\,\ty^3$ by means
of map $\tau^{-1}$. As result we get another frame $\tilde\bold E_1,\,
\tilde\bold E_2,\,\tilde\bold E_3$:
$$
\hskip -2em
\tilde\bold E_i=\sum^3_{r=1}\tilde T^r_i\,\bold E_s(\ty^1,\ty^2,\ty^3),
\qquad i=1,2,3.
\tag4.7
$$
Mutual scalar products of frame vectors \thetag{4.7} form a matrix
with components
$$
\hskip -2em
G_{ij}=(\tilde\bold E_i,\,\tilde\bold E_j)=
\sum^3_{r=1}\sum^3_{s=1}g_{rs}(\ty^1,\ty^2,\ty^3)\,\tilde T^r_i
\,\tilde T^s_j.
\tag4.8
$$
Upon transforming all arguments in \thetag{4.8} to $t,\,y^1,\,y^2,
\,y^3$ we get tensor field $\bold G$ with components $G_{ij}=G_{ij}
(t,\,y^1,\,y^2,\,y^3)$. Tensor $\bold G$ at the point with
coordinates $y^1,\,y^2,\,y^3$ is an exact quantitative measure of
deformation of our medium at that point. For small deformations
described in Cartesian coordinates we have
$$
\hskip -2em
G_{ij}(t,\,x^1,\,x^2,\,x^3)=g_{ij}-2\,u_{ij}\,+\,\ldots
\tag4.9
$$
Tensor $\bold u$ in \thetag{4.9} is standard {\it tensor of deformation}
as defined in \cite{4}:
$$
\hskip -2em
u_{ij}=\frac{1}{2}\left(\frac{\partial u_i}{\partial x^j}+
\frac{\partial u_j}{\partial x^i}\right).
\tag4.10
$$
By dots in \thetag{4.9} we denote terms of higher order with respect to
small displacements $u^1=\delta x^1,\,u^2=\delta x^2,\,u^3=\delta x^3$.
Relying upon above formulas \thetag{4.9} and \thetag{4.10}, now we define
deformation tensor $\bold u$ as follows:
$$
\hskip -2em
u_{ij}=\frac{g_{ij}-G_{ij}}{2}.
\tag4.11
$$
Using \thetag{4.11} instead of \thetag{4.10}, we have not to restrict
ourself to small displacements and can consider deformations of any
magnitude in any curvilinear coordinates.\par
    Let's calculate time derivative for deformation tensor $\bold u$
defined by formula \thetag{4.11}. Differentiating \thetag{4.8}, by
direct calculations we derive the following formula:
$$
\hskip -2em
\dot G_{ij}=-\sum^3_{k=1}\nabla_{\!k}G_{ij}\,v^k-\sum^3_{k=1}
G_{kj}\,\nabla_{\!i}v^k-\sum^3_{k=1}G_{ik}\,\nabla_{\!j}v^k.
\tag4.12
$$
Now, combining formulas \thetag{4.11} and \thetag{4.12}, we obtain
ultimate expression for $\dot u_{ij}$:
$$
\hskip -2em
\dot u_{ij}=\frac{\nabla_{\!i}v_j+\nabla_{\!j}v_i}{2}
-\sum^3_{k=1}\left(\nabla_{\!k}u_{ij}\,v^k+u_{kj}\,\nabla_{\!i}v^k
+u_{ik}\,\nabla_{\!j}v^k\right).
\tag4.13
$$
Let's denote by $v_{ij}$ first term in right hand side of \thetag{4.13}:
$$
\hskip -2em
v_{ij}=\frac{\nabla_{\!i}v_j+\nabla_{\!j}v_i}{2}.
\tag4.14
$$
Last three terms under summation in formula \thetag{4.13} are nonlinear
with respect to deformation functions \thetag{4.1}. They are usually
omitted in the case of small deformations. In that case we would have
$\dot u_{ij}=v_{ij}$.\par
\head
5. Dynamics of continuous medium.
\endhead
    Dynamics of any continuous medium (either liquid, solid, or gaseous)
is usually described in terms of three balance equations. These are
equations for 
\roster
\item mass balance;
\item momentum balance;
\item energy balance.
\endroster
Mass balance is most simple among balance equations. It is written on
the base of the following statement: change of mass enclosed in any
fixed volume within continuous medium is determined by mass flow through
its boundary:
$$
\hskip -2em
\frac{\partial\rho}{\partial t}
+\sum^3_{k=1}\nabla_{\!k}(\,\rho\,v^k)=0.
\tag5.1
$$
Sum in \thetag{5.1} is divergency $\divr\bold j$, where $\bold j
=\rho\bold v$ is density vector for mass flow. Components of velocity
vector are determined by formula \thetag{4.5}, while operator $\nabla_{\!k}$
in \thetag{5.1} should be applied according to the formula \thetag{3.2}.
\par
    Momentum balance equation is more complicated, though it has the
structure similar to mass balance equation \thetag{5.1}. Momentum balance
is written as
$$
\hskip -2em
\frac{\partial(\,\rho\,v^i)}{\partial t}
+\sum^3_{k=1}\nabla_{\!k}\Pi^{ik}=f^i.
\tag5.2
$$
Vector $\bold f$ with components $f^1,\,f^2,\,f^3$ in right hand side
of \thetag{5.2} determines density of volume forces in continuous medium.
Symmetric tensor $\boldsymbol\Pi$ with components $\Pi^{ik}$ determines
density of momentum flow in continuous medium. Exact formulas for
tensors $\bold f$ and $\bold\Pi$ depend on various properties of medium
and on those phenomena we are going to consider.\par
    Energy per unit of volume in continuous medium is a sum of two
components: {\it kinetic energy} and inner {\it thermal energy} due
to chaotic motion of atoms and molecules. Therefore energy balance
equation is written as follows:
$$
\hskip -2em
\frac{\partial}{\partial t}\!\left(\frac{\rho\,|\bold v|^2}{2}+
\rho\,\varepsilon\right)+\sum^3_{k=1}\nabla_{\!k}w^k=e.
\tag5.3
$$
Vector $\bold w$ with components $w^1,\,w^2,\,w^3$ in \thetag{5.3}
determines density of energy flow. Scalar $\bold e$ determines
energy production/dissipation per unit volume of medium. It is due
to work of force $\bold f$ in \thetag{5.2} and from other possible
sources (e\.\,g\. Joule heating due to electric current in conducting
medium). Potential energy is not included into left hand side of
\thetag{5.3}. However, work of potential forces is taken into account
among other terms as a part of scalar $e$ in right hand side of
\thetag{5.3}.\par
    Deformation state of a medium is completely determined by
tensor $\bold G$ and metric tensor $\bold g$. From \thetag{4.12}
and \thetag{5.1} one easily derives:
$$
\pagebreak
\hskip -2em
\ln\rho-\frac{\ln\det\bold G-\ln\det\bold g}{2}=\const.
\tag5.4
$$
Tensor $\boldsymbol\Pi$ in \thetag{5.2} is usually given by the following
formula:
$$
\hskip -2em
\Pi^{ik}=\rho\,v^i\,v^k-\sigma^{ik}.
\tag5.5
$$
Here $\sigma^{ik}$ is {\it stress tensor}. Substituting \thetag{5.5}
into \thetag{5.2}, we get
$$
\hskip -2em
\frac{\partial(\,\rho\,v^i)}{\partial t}
+\sum^3_{k=1}\nabla_{\!k}(\,\rho\,v^i\,v^k)=f^i+\sum^3_{k=1}
\nabla_{\!k}\sigma^{ik}.
\tag5.6
$$
Now, taking into account \thetag{5.1}, from \thetag{5.6} we derive
$$
\hskip -2em
\frac{\partial v^i}{\partial t}+\sum^3_{k=1}v^k\,\nabla_{\!k}v^i
=\frac{f^i}{\rho}+\sum^3_{k=1}\frac{\nabla_{\!k}\sigma^{ik}}{\rho}.
\tag5.7
$$
Using \thetag{5.4}, now we can calculate time derivative for the
density of kinetic energy:
$$
\hskip -2em
\frac{\partial}{\partial t}\!\left(\frac{\rho\,|\bold v|^2}{2}
\right)+\sum^3_{k=1}\nabla_{\!k}\!\left(\frac{\rho\,|\bold v|^2}{2}
\,v^k\!\right)=\sum^3_{i=1}v_i\,f^i+\sum^3_{i=1}\sum^3_{k=1}v_i\,
\nabla_{\!k}\sigma^{ik}.
\tag5.8
$$
Looking at formula \thetag{5.8}, we see that right hand side of
this formula is completely determined by parameters $\sigma^{ik}$
and $f^i$ from \thetag{5.5} and \thetag{5.2}.
\head
6. Elastic solids state media.
\endhead
    Deformation state of a medium is completely determined by
tensor $\bold G$ and metric tensor $\bold g$. Therefore $\varepsilon$
in \thetag{5.3} satisfies the following equality:
$$
\hskip -2em
d\varepsilon=T\,ds-\sum^3_{i=1}\sum^3_{j=1}\frac{\bar\sigma^{ij}\,
dG_{ij}}{2\,\rho}.
\tag6.1
$$
Here $s$ is an entropy per unit mass of solid state medium. Note that
$\bar\sigma^{ij}$ in \thetag{6.1} should not coincide with $\sigma^{ij}$
considered above. Free energy per unit mass is determined by standard
formula $f=\varepsilon-T\,s$. Therefore
$$
\hskip -2em
df=-s\,dT-\sum^3_{i=1}\sum^3_{j=1}\frac{\bar\sigma^{ij}\,dG_{ij}}{2\,\rho}.
\tag6.2
$$
Free energy per unit mass $f$ is a function of temperature $T$ and
deformation state of medium: $f=f(T,\bold G)$. From \thetag{6.2} we
derive
$$
\hskip -2em
\frac{\bar\sigma^{ij}}{2\,\rho}=-\frac{\partial f(T,\bold G)}
{\partial G_{ij}}.
\tag6.3
$$
Solid materials can exhibit different properties in different
directions. We consider only those polymer materials, \pagebreak
which are homogeneous and isotropic (i\.\,e\. uniform in all
directions). For such materials $f(T,\bold G)$ is given by formula
$$
\hskip -2em
f=f(T,\lambda_{[1]},\lambda_{[2]},\lambda_{[3]}),
\tag6.4
$$
where $\lambda_{[1]},\,\lambda_{[2]},\,\lambda_{[3]}$ are three
scalar invariants for linear operator $\bold G$:
$$
\xalignat 3
&\lambda_{[1]}=\frac{\tr(\bold G)}{3},
&&\lambda_{[2]}=\frac{\tr(\bold G\cdot\bold G)}{3},
&&\lambda_{[3]}=\frac{\tr(\bold G\cdot\bold G\cdot\bold G)}{3}.
\qquad
\tag6.5
\endxalignat
$$
Linear operator $\bold G$ in \thetag{6.5} is determined by its
matrix $G^{\,i}_j$, where
$$
\hskip -2em
G^{\,i}_j=\sum^3_{k=1}g^{ik}\,G_{kj}.
\tag6.6
$$
Using \thetag{6.5} and \thetag{6.6}, we calculate partial derivatives
$$
\xalignat 3
&\frac{\partial\lambda_{[1]}}{\partial G_{ij}}=\frac{g^{ij}}{3},
&&\frac{\partial\lambda_{[2]}}{\partial G_{ij}}=\frac{2\,G^{ij}}{3},
&&\frac{\partial\lambda_{[3]}}{\partial G_{ij}}=\sum^3_{k=1}
\sum^3_{q=1}G^{ik}\,g_{kq}\,G^{qj}.
\qquad
\tag6.7
\endxalignat
$$
Applying \thetag{6.3} and \thetag{6.7} to \thetag{6.4}, we find
most general formula for $\bar\sigma^{ij}$:
$$
\hskip -2em
\bar\sigma^{ij}=f_{[1]}\ g^{ij}+f_{[2]}\ G^{ij}+\sum^3_{k=1}
\sum^3_{q=1}f_{[3]}\ G^{ik}\,g_{kq}\,G^{qj},
\tag6.8
$$
Here $f_{[1]},\,f_{[2]},\,f_{[3]}$ are coefficients depending on
$T$ and on scalar invariants \thetag{6.5}:
$$
\hskip -2em
f_{[i]}=-\frac{2\,i\,\rho}{3}\,\frac{\partial f(T,\lambda_{[1]},
\lambda_{[2]},\lambda_{[3]})}{\partial\lambda_{[i]}},
\qquad i=1,2,3.
\tag6.9
$$
Formula \thetag{6.8} is exact, but very huge. Its use in numeric
simulation is impossible not because of huge computations, but
since function \thetag{6.4} is not properly measured experimentally
for broad range of its arguments. Formulas \thetag{6.8} and
\thetag{6.9} are worth for us only because they indicate the dependence
$$
\hskip -2em
\bar\sigma^{ij}=\bar\sigma^{ij}(T,\bold G).
\tag6.10
$$
Traditionally linearized version of this dependence \thetag{6.10}
is used for the case of small deformations when $u_{ij}$ are much
less than $g_{ij}$.\par
    Now let's consider formula \thetag{6.2} again. If function \thetag{6.4}
is known, entropy per unit mass can be calculated as partial derivative:
$$
\hskip -2em
s=-\frac{\partial f}{\partial T}.
\tag6.11
$$
Due to \thetag{6.11} we can treat $s$ as a function $s=s(T,\bold G)$.
Moreover, $s$ depends on $\bold G$ through scalar invariants
\thetag{6.5} just like function $f$ itself:
$$
s=s(T,\lambda_{[1]},\lambda_{[2]},\lambda_{[3]}).
\tag6.12
$$
Then for thermal energy per unit mass we get
$$
\hskip -2em
\varepsilon=f+T\,s=\varepsilon(T,\bold G)=
\varepsilon(T,\lambda_{[1]},\lambda_{[2]},\lambda_{[3]}).
\tag6.13
$$
If parameters $w^k$ and $e$ in \thetag{5.3} are known, then, substituting
\thetag{6.13} into \thetag{5.3}, we get the equation for temperature
function $T=T(t,y^1,y^2,y^3)$.\par
    In thermodynamics specific thermal energy $\varepsilon$ is often
treated as a function of entropy. Indeed, inverting the dependence of
$s$ on $T$ for fixed $\lambda_{[1]},\,\lambda_{[2]},\,\lambda_{[3]}$
in \thetag{6.12} and substituting $T=T(s,\lambda_{[1]},\lambda_{[2]},
\lambda_{[3]})$ into \thetag{6.13}, we get
$$
\hskip -2em
\varepsilon=\varepsilon(s,\bold G)=
\varepsilon(s,\lambda_{[1]},\lambda_{[2]},\lambda_{[3]}).
\tag6.14
$$
In the absence of heat transfer and viscosity, parameters $w^k$ and
$e$ can be derived from entropy balance equation. In this case dynamics
of solid state medium is {\it adiabatic}. Therefore one can write the
equation
$$
\hskip -2em
\frac{\partial(\,\rho\,s)}{\partial t}
+\sum^3_{k=1}\nabla_{\!k}(\,\rho\,s\,v^k)=0.
\tag6.15
$$
By analogy with \thetag{6.15}, now we calculate the following quantity:
$$
\hskip -2em
\frac{\partial(\,\rho\,\varepsilon)}{\partial t}
+\sum^3_{k=1}\nabla_{\!k}(\,\rho\,\varepsilon\,v^k)=
\rho\,\frac{\partial\varepsilon}{\partial t}
+\sum^3_{k=1}\rho\,v^k\,\nabla_{\!k}\varepsilon.
\tag6.16
$$
Applying \thetag{6.1} and \thetag{6.14} and using $\nabla_{\!k}g_{ij}=0$,
which is basic property of metric tensor, we get formulas for $\partial
\varepsilon/\partial t$ and $\nabla_{\!k}\varepsilon$ in right hand side
of \thetag{6.16}:
$$
\hskip -2em
\aligned
&\frac{\partial\varepsilon}{\partial t}=T\,\frac{\partial s}{\partial t}
-\sum^3_{i=1}\sum^3_{j=1}\frac{\bar\sigma^{ij}\,
\dot G_{ij}}{2\,\rho},\\
&\nabla_{\!k}\varepsilon=T\,\nabla_{\!k}s-\sum^3_{i=1}\sum^3_{j=1}
\frac{\bar\sigma^{ij}\,\nabla_{\!k}G_{ij}}{2\,\rho}.
\endaligned
\tag6.17
$$
Substituting \thetag{6.17} into \thetag{6.16} and taking into
account \thetag{6.15} and \thetag{5.1}, we get
$$
\hskip -2em
\frac{\partial(\,\rho\,\varepsilon)}{\partial t}
+\sum^3_{k=1}\nabla_{\!k}(\,\rho\,\varepsilon\,v^k)=
-\sum^3_{i=1}\sum^3_{j=1}\frac{\bar\sigma^{ij}}{2}
\left(\!\dot G_{ij}+\shave{\sum^3_{k=1}}v^k\,\nabla_{\!k}
G_{ij}\!\right).
\tag6.18
$$
Now let's substitute \thetag{4.12} into \thetag{6.18}. As a result
of simple calculations we find
$$
\gather
\frac{\partial(\,\rho\,\varepsilon)}{\partial t}
+\sum^3_{k=1}\nabla_{\!k}(\,\rho\,\varepsilon\,v^k)=
\sum^3_{i=1}\sum^3_{j=1}\frac{\bar\sigma^{ij}}{2}
\left(\,\shave{\sum^3_{k=1}}\nabla_{\!j}v^k\,G_{ik}+
\shave{\sum^3_{k=1}}\nabla_{\!i}v^k\,G_{kj}\!\right)=\\
=\sum^3_{i=1}\sum^3_{j=1}\sum^3_{k=1}\frac{\bar\sigma^{ik}
\,G^{\,j}_k+\bar\sigma^{ki}\,G^{\,j}_k}{2}\,\nabla_{\!i}v_j
=\sum^3_{i=1}\sum^3_{k=1}\nabla_{\!k}v_i\left(\,
\shave{\sum^3_{j=1}}G^{\,i}_j\,\bar\sigma^{jk}\right)\!.
\endgather
$$
Note that the above formula is quite similar to formula \thetag{5.8}.
Adding these two formulas, we obtain the following equality:
$$
\hskip -2em
\gathered
\frac{\partial}{\partial t}\!\left(\frac{\rho\,|\bold v|^2}{2}
+\rho\,\varepsilon\right)+\sum^3_{k=1}\nabla_{\!k}\!\left(
\frac{\rho\,|\bold v|^2}{2}\,v^k+\rho\,\varepsilon
\,v^k\!\right)=\\
=\sum^3_{i=1}v_i\,f^i+\sum^3_{i=1}\sum^3_{k=1}v_i\,
\nabla_{\!k}\sigma^{ik}+\sum^3_{i=1}\sum^3_{k=1}
\nabla_{\!k}v_i\left(\,\shave{\sum^3_{j=1}}G^{\,i}_j\,
\bar\sigma^{jk}\right)\!.
\endgathered
\tag6.19
$$
Right hand side of the equality \thetag{6.19} simplifies
substantially if stress tensor $\sigma^{ik}$ and tensor
$\bar\sigma^{ij}$ introduced in \thetag{6.1} are related
as follows:
$$
\hskip -2em
\sigma^{ik}=\sum^3_{j=1}G^{\,i}_j\,\bar\sigma^{jk}.
\tag6.20
$$
Due to the relationship \thetag{6.20} formula \thetag{6.19}
transforms to the following one:
$$
\hskip -2em
\gathered
\frac{\partial}{\partial t}\!\left(\frac{\rho\,|\bold v|^2}{2}
+\rho\,\varepsilon\right)+\sum^3_{k=1}\nabla_{\!k}\!\left(
\frac{\rho\,|\bold v|^2}{2}\,v^k+\rho\,\varepsilon
\,v^k\!\right)=\\
=\sum^3_{i=1}v_i\,f^i+\sum^3_{i=1}\sum^3_{k=1}
\nabla_{\!k}(v_i\,\sigma^{ik}).
\endgathered
\tag6.21
$$
The relationship \thetag{6.20} is self consistent. Indeed, if
\thetag{6.20} is fulfilled then we have the equality \thetag{6.21}
with quite transparent interpretation. First two terms in
\thetag{6.21} describe energy increment and energy flow due to
mass transport. Two terms in right hand side of \thetag{6.21}
describe energy creation due to external forces and due to stress
forces in continuous medium. Comparing \thetag{6.21} and \thetag{5.3},
we get
$$
\xalignat 2
&w^k=\frac{\rho\,|\bold v|^2}{2}\,v^k+\rho\,\varepsilon\,v^k
-\sum^3_{i=1}v_i\,\sigma^{ik},
&&e=\sum^3_{i=1}v_i\,f^i.
\qquad
\tag6.22
\endxalignat
$$
Due to special form of tensor $\bar{\boldsymbol\sigma}$ and due
to \thetag{6.20} stress tensor $\boldsymbol\sigma$ is symmetric:
$\sigma^{ij}=\sigma^{ji}$. Indeed, from formula \thetag{6.8} for
components of stress tensor we derive
$$
\hskip -2em
\gathered
\sigma^{ij}=f_{[1]}\ G^{ij}+\sum^3_{k=1}\sum^3_{q=1}f_{[2]}\
G^{ik}\,g_{kq}\,G^{qj}\,+\\
+\,\sum^3_{k=1}\sum^3_{q=1}\sum^3_{m=1}\sum^3_{n=1}f_{[3]}\
G^{ik}\,g_{kq}\,G^{qm}\,g_{mn}\,G^{nj}.
\endgathered
\tag6.23
$$
Remember that coefficients $f_{[1]},\,f_{[2]},\,f_{[3]}$ in
\thetag{6.23} are determined by formula \thetag{6.9}. They
depend on deformation $\bold G$ and temperature $T$.\par
\head
7. Heat transfer, viscosity, and entropy production.
\endhead
    Formulas for $\Pi^{ik}$ and $w^k$ in balance equations
\thetag{5.2} and \thetag{5.3} become more complicated \pagebreak
if we take into account viscosity and thermal conductivity of
medium. In the case when $\nabla\bold v\neq 0$ different parts
of medium immediately adjacent to each other move with different
velocities. This gives rise to forces of viscous friction.
These forces are described by additional term in \thetag{5.5}:
$$
\hskip -2em
\Pi^{ik}=\rho\,v^i\,v^k-\sigma^{ik}-\tilde\sigma^{ik}.
\tag7.1
$$
Here $\tilde\sigma^{ik}$ are components of {\it viscous stress
tensor} $\tilde{\boldsymbol\sigma}$. In linear approximation they
are linear with respect to velocity gradients:
$$
\hskip -2em
\tilde\sigma^{ik}=\sum^3_{j=1}\sum^3_{q=1}\eta^{ikjq}\,v_{jq}.
\tag7.2
$$
Here $v_{jq}$ are components of symmetric tensor introduced in
\thetag{4.14}, while $\eta^{ikjq}$ in \thetag{7.2} are components
of {\it viscosity tensor}. They possess the following symmetry:
$$
\hskip -2em
\eta^{ikjq}=\eta^{jqik}=\eta^{kijq}=\eta^{ikqj}.
\tag7.3
$$
Components of viscosity tensor \thetag{7.3} are kinetic coefficients.
In near equilibrium deformations they are functions of temperature
and deformation tensor:
$$
\hskip -2em
\eta^{ikjq}=\eta^{ikjq}(T,\bold G).
\tag7.4
$$\par
    If continuous medium is non-uniformly heated, i\.\,e\. $\nabla T
\neq 0$, this may cause direct heat transfer without mass transport
in it. This phenomenon is due to {\it thermal conductivity} of medium.
Thermal conductivity of medium is described by additional term in
formula for density of energy flow $\bold w$:
$$
\hskip -2em
w^k=\frac{\rho\,|\bold v|^2}{2}\,v^k+\rho\,\varepsilon\,v^k
-\sum^3_{i=1}v_i\,\sigma^{ik}-\sum^3_{i=1}v_i\,\tilde\sigma^{ik}
-\sum^3_{i=1}\nabla_{\!i}T\,\varkappa^{ik}.
\tag7.5
$$
As compared to \thetag{6.22}, in \thetag{7.5} we have two extra terms.
First is due to viscous stress tensor $\tilde{\boldsymbol\sigma}$,
second term contains components of {\it heat conductivity tensor}
$\varkappa^{ik}$. Like $\eta^{ikjq}$ in \thetag{7.4}, components of
heat conductivity tensor are kinetic coefficients, they depend on
temperature $T$ and deformation $\bold G$:
$$
\hskip -2em
\varkappa^{ik}=\varkappa^{ik}(T,\bold G).
\tag7.6
$$
Heat conductivity tensor \thetag{7.6} is symmetric, i\.\,e\.
$\varkappa^{ik}=\varkappa^{ki}$.\par
    Let's recalculate entropy balance equation \thetag{6.15}, taking
into account additional terms in \thetag{7.1} and \thetag{7.5}.
Instead of \thetag{6.21} now we have
$$
\hskip -2em
\gathered
\frac{\partial}{\partial t}\!\left(\frac{\rho\,|\bold v|^2}{2}
+\rho\,\varepsilon\right)+\sum^3_{k=1}\nabla_{\!k}\!\left(
\frac{\rho\,|\bold v|^2}{2}\,v^k+\rho\,\varepsilon
\,v^k\!\right)=\\
=\sum^3_{i=1}v_i\,f^i+\sum^3_{i=1}\sum^3_{k=1}
\nabla_{\!k}(v_i\,\sigma^{ik}+v_i\,\tilde\sigma^{ik})
+\sum^3_{i=1}\sum^3_{k=1}\nabla_{\!k}(\nabla_{\!i}T\,
\varkappa^{ik}).
\endgathered
\tag7.7
$$
Equations \thetag{5.7} and \thetag{5.8} are replaced by the
following two equalities respectively:
$$
\align
&\hskip -2em
\frac{\partial v^i}{\partial t}+\sum^3_{k=1}v^k\,\nabla_{\!k}v^i
=\frac{f^i}{\rho}+\sum^3_{k=1}\frac{\nabla_{\!k}\sigma^{ik}}{\rho}
+\sum^3_{k=1}\frac{\nabla_{\!k}\tilde\sigma^{ik}}{\rho},\\
\vspace{2ex}
&\hskip -2em
\aligned
\frac{\partial}{\partial t}\!\left(\frac{\rho\,|\bold v|^2}{2}
\right)&+\sum^3_{k=1}\nabla_{\!k}\!\left(\frac{\rho\,|\bold v|^2}{2}
\,v^k\!\right)=\\
&\hskip -4em=\sum^3_{i=1}v_i\,f^i+\sum^3_{i=1}\sum^3_{k=1}v_i
\,\nabla_{\!k}\sigma^{ik}+\sum^3_{i=1}\sum^3_{k=1}v_i\,\nabla_{\!k}
\tilde\sigma^{ik}.
\endaligned
\tag7.8
\endalign
$$
Subtracting \thetag{7.8} from \thetag{7.7} and taking into account
\thetag{5.1}, we obtain
$$
\rho\,\frac{\partial\varepsilon}{\partial t}
+\sum^3_{k=1}\rho\,v^k\,\nabla_{\!k}\varepsilon=
\sum^3_{i=1}\sum^3_{k=1}\left(\nabla_{\!k}v_i\,(\sigma^{ik}
+\tilde\sigma^{ik})+\nabla_{\!k}(\nabla_{\!i}T\,
\varkappa^{ik})\right).
\quad
\tag7.9
$$
For derivatives $\partial\varepsilon/\partial t$ and $\nabla_{\!k}
\varepsilon$ in left hand side of \thetag{7.9} we can apply
\thetag{6.17}. Then
$$
\gathered
T\,\frac{\partial(\,\rho\,s)}{\partial t}+\sum^3_{k=1}T\,
\nabla_{\!k}(\,\rho\,s\,v^k)-\sum^3_{i=1}\sum^3_{j=1}
\frac{\bar\sigma^{ij}}{2}\left(\!\dot G_{ij}+\shave{\sum^3_{k=1}}
v^k\,\nabla_{\!k}G_{ij}\!\right)=\\
=\sum^3_{i=1}\sum^3_{k=1}\left(\nabla_{\!k}v_i\,(\sigma^{ik}
+\tilde\sigma^{ik})+\nabla_{\!k}(\nabla_{\!i}T\,
\varkappa^{ik})\right).
\endgathered
\quad
\tag7.10
$$
Now let's substitute \thetag{4.12} into \thetag{7.10} and take
into account formula \thetag{6.20}:
$$
\gathered
\frac{\partial(\,\rho\,s)}{\partial t}+\sum^3_{k=1}\nabla_{\!k}
(\,\rho\,s\,v^k)=\sum^3_{i=1}\sum^3_{k=1}\frac{\nabla_{\!k}v_i\,
\tilde\sigma^{ik}+\nabla_{\!k}(\nabla_{\!i}T\,\varkappa^{ik})}
{T}=\\
=\sum^3_{i=1}\sum^3_{k=1}\nabla_{\!k}\!\left(\frac{\nabla_{\!i}T
\,\varkappa^{ik}}{T}\right)+
\sum^3_{i=1}\sum^3_{k=1}\sum^3_{j=1}\sum^3_{q=1}\frac{v_{ik}\,
\eta^{ikjq}\,v_{jq}}{T}+\sum^3_{i=1}\sum^3_{k=1}\frac{\nabla_{\!i}T
\,\varkappa^{ik}\,\nabla_{\!k}T}{T^2}.
\endgathered
$$
In ultimate form this is entropy balance equation generalizing
equation \thetag{6.15}:
$$
\hskip -2em
\gathered
\frac{\partial(\,\rho\,s)}{\partial t}+\sum^3_{k=1}\nabla_{\!k}\!
\left(\rho\,s\,v^k-\shave{\sum^3_{i=1}}\frac{\nabla_{\!i}T\,
\varkappa^{ik}}{T}\right)=\\
=\sum^3_{i=1}\sum^3_{k=1}\sum^3_{j=1}\sum^3_{q=1}\frac{v_{ik}\,
\eta^{ikjq}\,v_{jq}}{T}+\sum^3_{i=1}\sum^3_{k=1}\frac{\nabla_{\!i}T
\,\varkappa^{ik}\,\nabla_{\!k}T}{T^2}.
\endgathered
\tag7.11
$$
Last two terms in \thetag{7.11} are positive. This means that
total entropy is ever growing, when medium is evolving toward
thermodynamic equilibrium.\par
\head
8. Liquid state media.
\endhead
    Balance equations in liquids are quite similar to those in
elastic solid materials. The only difference is that liquids are
always isotropic. Deformation state of liquid material is completely
determined by its density $\rho$. Therefore, instead of \thetag{6.13},
for specific thermal energy per unit mass we have
$$
\hskip -2em
\varepsilon=\varepsilon(T,\rho).
\tag8.1
$$
Like $\varepsilon$ in \thetag{8.1}, specific free energy per unit mass
is given by formula
$$
\hskip -2em
f=f(T,\rho).
\tag8.2
$$
Density $\rho$ is related to deformation $\bold G$ by means of formula
\thetag{5.4}. Hence
$$
\hskip -2em
\frac{d\rho}{\rho}=\frac{d(\ln\det\bold G)}{2}=\frac{\tr(\bold G^{-1}
\,d\bold G)}{2}=\sum^3_{i=1}\sum^3_{j=1}\frac{\bar G^{ij}\,dG_{ij}}{2},
\tag8.3
$$
where $\bar G^{ij}$ are components of inverse matrix $\bold G^{-1}$.
Differentiating \thetag{8.2} and taking into account \thetag{8.3},
due to the equality \thetag{6.2} we have
$$
\hskip -2em
\bar\sigma^{ij}=-\rho^2\,\frac{\partial f}{\partial\rho}\,\bar G^{ij}.
\tag8.4
$$
Substituting \thetag{8.4} into \thetag{6.20}, we obtain formula for
stress tensor in liquid medium:
$$
\hskip -2em
\sigma^{ij}=-\rho^2\,\frac{\partial f}{\partial\rho}\,g^{ij}.
\tag8.5
$$
Scalar factor in \thetag{8.5} is interpreted as {\it pressure}.
Indeed, we have
$$
\xalignat 2
&\hskip -2em
p=\rho^2\,\frac{\partial f}{\partial\rho},
&&\sigma^{ij}=-p\,g^{ij}.
\tag8.6
\endxalignat
$$
Viscosity tensor in liquids simplifies and takes the following form:
$$
\hskip -2em
\eta^{ikjq}=\eta\,\left(\,g^{ij}\,g^{kq}+g^{iq}\,g^{jk}\right)+
\left(\!\zeta-\frac{2}{3}\,\eta\!\right)\,g^{ik}\,g^{jq}.
\tag8.7
$$
Heat conductivity tensor in liquid medium is also simpler than in
solid media:
$$
\hskip -2em
\varkappa^{ik}=\varkappa\,g^{ik}.
\tag8.8
$$
Pressure $p$ in \thetag{8.6} and scalar parameters $\eta$, $\zeta$,
and $\varkappa$ in \thetag{8.7} and \thetag{8.8} all are functions
of temperature $T$ and density $\rho$ of continuous medium:
$$
\xalignat 2
&\hskip -2em
p=p(T,\rho),&&\eta=\eta(T,\rho),\\
&\hskip -2em
\zeta=\zeta(T,\rho),&&\varkappa=\varkappa(T,\rho).
\endxalignat
$$
Other parameters in balance equations \thetag{5.1}, \thetag{5.2},
\thetag{5.3} for liquids are same as for solid state media.\par
\head
9. Plastic materials.
\endhead
    Saying plastic materials we mean pitch-like very dense sticky
liquids and many solid materials with no crystalline grid, e\.\,g\.
glass and polymer materials. They resist to deformations like
solids and can flow like liquids, though sometimes very slowly. In
order to describe such materials mathematically we need to divide
deformation $\bold G$ into two parts: elastic deformation
$\hat{\bold G}$ and plastic deformation $\check{\bold G}$:
$$
\hskip -2em
G_{ij}=\sum^3_{k=1}\sum^3_{q=1}\check G^{\,k}_i\,\hat G_{kq}
\,\check G^{\,q}_j.
\tag9.1
$$
Like $G_{ij}$, {\it elastic deformation tensor} $\hat G_{ik}$ in
\thetag{9.1} is symmetric: $\hat G_{ik}=\hat G_{ki}$. Plastic
deformation tensor $\check G_{kj}$ is not necessarily symmetric.
\par
    Plastic deformation arises as a response to stress tending to
relax this stress. Elastic deformation tensor $\hat{\bold G}$ is
thermodynamic parameter of plastic medium, while plastic deformation
tensor $\check{\bold G}$ is kinetic parameter. Therefore all
thermodynamic quantities and kinetic coefficients for near thermodynamic
equilibrium deformations of plastic medium depend on elastic deformation
tensor $\hat{\bold G}$ and on temperature:
$$
\xalignat 2
&\hskip -2em
\varepsilon=\varepsilon(T,\hat{\bold G}),
&&f=f(T,\hat{\bold G}),\\
&\hskip -2em
\bar\sigma^{ij}=\bar\sigma^{ij}(T,\hat{\bold G}),
&&\sigma^{ij}=\sigma^{ij}(T,\hat{\bold G}),
\tag9.2\\
&\hskip -2em
\eta^{ikjq}=\eta^{ikjq}(T,\hat{\bold G}),
&&\varkappa^{ik}=\varkappa^{ik}(T,\hat{\bold G}).
\endxalignat
$$
Components of total deformation tensor $G_{ij}$ in \thetag{9.1} satisfy
differential equations \thetag{4.12}. These equations can be written as
follows:
$$
\hskip -2em
\frac{\partial G_{ij}}{\partial t}+\sum^3_{r=1}v^r\,\nabla_{\!r}G_{ij}
=-\sum^3_{r=1}\nabla_{\!i}v^r\,G_{rj}-\sum^3_{r=1}G_{ir}\,\nabla_{\!j}v^r.
\tag9.3
$$
For components of elastic deformation tensor $\hat G_{ij}$ we write
analogous equation:
$$
\hskip -2em
\frac{\partial\hat G_{kq}}{\partial t}+\sum^3_{r=1}v^r\,\nabla_{\!r}
\hat G_{kq}=-\sum^3_{r=1}\nabla_{\!k}v^r\,\hat G_{rq}-\sum^3_{r=1}
\hat G_{kr}\,\nabla_{\!q}v^r+2\,\Theta_{kq}.
\tag9.4
$$
Components of symmetric tensor $\boldsymbol\Theta$ in \thetag{9.4} are
kinetic coefficients. Therefore
$$
\hskip -2em
\Theta_{kq}=\Theta_{kq}(T,\hat{\bold G}).
\tag9.5
$$
Now we need some special facts concerning all three deformation
tensors in \thetag{9.1}. Symmetric matrix $G_{ij}$ determined by
formula \thetag{4.8} is non-degenerate and positive. Due to
\thetag{9.1} we have the equality $\det\bold G=\det\hat{\bold G}
\cdot(\det\check{\bold G})^2$. Hence
$$
\xalignat 2
&\hskip -2em
\det\hat{\bold G}\neq 0,
&&\det\check{\bold G}\neq 0.
\tag9.6
\endxalignat
$$
Moreover, due to \thetag{9.1} and \thetag{9.6} matrix $\hat G_{kq}$
\pagebreak is also positive. Now let's use formula analogous to
\thetag{6.6} and define linear operator $\hat{\bold G}$ with components
$$
\hskip -2em
\hat G^{\,i}_j=\sum^3_{k=1}g^{ik}\,\hat G_{kj}.
\tag9.7
$$
\proclaim{Theorem 9.1} For any symmetric matrix $\Theta_{kq}$
and for linear operator $\hat{\bold G}$ determined by formula
\thetag{9.7} there is unique symmetric matrix $\theta_{kq}$
such that
$$
\hskip -2em
2\,\Theta_{kq}=\sum^3_{r=1}\theta_{kr}\,\hat G^{\,r}_q
+\sum^3_{r=1}\hat G^{\,r}_k\,\theta_{rq}.
\tag9.8
$$
\endproclaim
This is purely mathematical fact with rather simple proof. Applying
theorem~9.1 to matrix \thetag{9.5} we get tensor $\boldsymbol\theta$
with components
$$
\hskip -2em
\theta_{kq}=\theta_{kq}(T,\hat{\bold G}).
\tag9.9
$$
Raising index in \thetag{9.9} we get symmetric linear operator
$\boldsymbol\theta$ with components
$$
\hskip -2em
\theta^{\,i}_j=\sum^3_{k=1}g^{ik}\,\theta_{kj}.
\tag9.10
$$
Due to \thetag{9.8}, \thetag{9.9}, and \thetag{9.10} we can write
equation \thetag{9.4} in the following form:
$$
\hskip -2em
\gathered
\frac{\partial\hat G_{kq}}{\partial t}+\sum^3_{r=1}v^r\,
\nabla_{\!r}\hat G_{kq}=-\sum^3_{r=1}\nabla_{\!k}v^r\,
\hat G_{rq}\,-\\
-\sum^3_{r=1}\hat G_{kr}\,\nabla_{\!q}v^r
+\sum^3_{r=1}
\theta^{\,r}_k\,\hat G_{rq}+\sum^3_{r=1}\hat G_{kr}
\,\theta^{\,r}_q.
\endgathered
\tag9.11
$$
Now we are able to write differential equations for plastic deformation
tensor:
$$
\hskip -2em
\frac{\partial\check G^{\,k}_i}{\partial t}+
\sum^3_{r=1}v^r\,\nabla_{\!r}\check G^{\,k}_i=
\sum^3_{r=1}\left(\check G^{\,r}_i\,\nabla_{\!r}v^k
-\nabla_{\!i}v^r\,\check G^{\,k}_r\right)-\sum^3_{r=1}
\theta^{\,k}_r\,\check G^{\,r}_i.
\tag9.12
$$\par
     {\bf Forgetting principle}. This is basic principle characterizing
plastic deformations that we consider in present paper. Suppose that
plastic medium evolves from initial state with no deformation at time
instant $t=0$ to some intermediate state at $t=t_0$, then it continues
its evolution for $t>t_0$. Forgetting principle states that if in
intermediate state total deformation of medium is purely plastic, then
further evolution of medium will be so as if in intermediate state it
had no deformation at all. Equation \thetag{9.12} is written on the base
of this principle. It is compatible with \thetag{9.1} and with equations
\thetag{9.3} and \thetag{9.11}.\par
\head
10. Thermodynamics of plastic medium.
\endhead
    Balance equations \thetag{5.1}, \thetag{5.2}, and \thetag{5.3}
remain unchanged for plastic medium. We also keep unchanged formulas
\thetag{7.1}, \thetag{7.2}, \thetag{7.5}, and second formula in
\thetag{6.22}. However, due to \thetag{9.2} and \thetag{9.11} we
should revise formulas \thetag{6.2}, \thetag{6.3}, \thetag{6.4},
\thetag{6.5}, and \thetag{6.6}. Formulas \thetag{6.2} and \thetag{6.3}
are replaced by the following ones:
$$
\xalignat 2
&\hskip -2em
df=-s\,dT-\sum^3_{i=1}\sum^3_{j=1}\frac{\bar\sigma^{ij}\,d\hat G_{ij}}
{2\,\rho},
&&\frac{\bar\sigma^{ij}}{2\,\rho}=-\frac{\partial f(T,\hat{\bold G})}
{\partial\hat G_{ij}}.
\quad
\tag10.1
\endxalignat
$$
For homogeneous and isotropic plastic material function $f(T,\hat{\bold
G})$ is more special:
$$
\hskip -2em
f=f(T,\lambda_{[1]},\lambda_{[2]},\lambda_{[3]}).
\tag10.2
$$
Though $\lambda_{[1]},\,\lambda_{[2]},\,\lambda_{[3]}$ in \thetag{10.2}
are different from that of \thetag{6.4}, they are scalar invariants of
linear operator $\hat{\bold G}$ that was defined above by formula
\thetag{9.7}:
$$
\xalignat 3
&\lambda_{[1]}=\frac{\tr(\hat{\bold G})}{3},
&&\lambda_{[2]}=\frac{\tr(\hat{\bold G}\cdot\hat{\bold G})}{3},
&&\lambda_{[3]}=\frac{\tr(\hat{\bold G}\cdot\hat{\bold G}\cdot
\hat{\bold G})}{3}.
\qquad
\tag10.3
\endxalignat
$$
Next two formulas are analogous to \thetag{6.8} and \thetag{6.9}:
$$
\gather
\hskip -2em
\bar\sigma^{ij}=f_{[1]}\ g^{ij}+f_{[2]}\ \hat G^{ij}+\sum^3_{k=1}
\sum^3_{q=1}f_{[3]}\ \hat G^{ik}\,g_{kq}\,\hat G^{qj},
\tag10.4\\
\vspace{2ex}
\hskip -2em
f_{[i]}=-\frac{2\,i\,\rho}{3}\,\frac{\partial f(T,\lambda_{[1]},
\lambda_{[2]},\lambda_{[3]})}{\partial\lambda_{[i]}},
\qquad i=1,2,3.
\tag10.5
\endgather
$$\par
   As we said above, we keep unchanged energy balance equation
\thetag{5.3} with $w^k$ given by formula \thetag{7.5} and $e$
given by second formula \thetag{6.22}. This means that we keep
unchanged formula \thetag{7.7}. However, thermal energy
$\varepsilon$ now depends only on elastic part of deformation
tensor. Therefore, we should recalculate relation of $\bar\sigma^{ij}$
and stress tensor $\sigma^{ij}$. Due to \thetag{10.1} and consequent
formulas \thetag{10.2}, \thetag{10.3}, \thetag{10.4}, and \thetag{10.5}
instead of \thetag{6.14} now we should write
$$
\hskip -2em
\varepsilon=\varepsilon(s,\hat{\bold G})=
\varepsilon(s,\lambda_{[1]},\lambda_{[2]},\lambda_{[3]}).
\tag10.6
$$
By the same reason instead of formulas \thetag{6.17} we should write
$$
\hskip -2em
\aligned
&\frac{\partial\varepsilon}{\partial t}=T\,\frac{\partial s}{\partial t}
-\sum^3_{i=1}\sum^3_{j=1}\frac{\bar\sigma^{ij}}{2\,\rho}\,
\frac{\partial\hat G_{ij}}{\partial t},\\
\vspace{2ex}
&\nabla_{\!k}\varepsilon=T\,\nabla_{\!k}s-\sum^3_{i=1}\sum^3_{j=1}
\frac{\bar\sigma^{ij}\,\nabla_{\!k}\hat G_{ij}}{2\,\rho}.
\endaligned
\tag10.7
$$
Applying \thetag{5.1} and \thetag{7.8} to \thetag{7.7}, we derive
the following equation:
$$
\rho\,\frac{\partial\varepsilon}{\partial t}
+\sum^3_{k=1}\rho\,v^k\,\nabla_{\!k}\varepsilon=
\sum^3_{i=1}\sum^3_{k=1}\left(\nabla_{\!k}v_i\,(\sigma^{ik}
+\tilde\sigma^{ik})+\nabla_{\!k}(\nabla_{\!i}T\,
\varkappa^{ik})\right).
\quad
\tag10.8
$$
Note that \thetag{10.8} is just the same as the equation \thetag{7.9}.
Now we should substitute \thetag{10.7} into this equation. As a result
we obtain
$$
\gathered
\rho\,T\,\frac{\partial s}{\partial t}
+\sum^3_{k=1}\rho\,T\,v^k\,\nabla_{\!k}s
-\sum^3_{i=1}\sum^3_{j=1}\frac{\bar\sigma^{ij}}{2}\!\left(\!
\frac{\partial\hat G_{ij}}{\partial t}+
\shave{\sum^3_{k=1}}v^k\,\nabla_{\!k}\hat G_{ij}\!\right)=
\\
=\sum^3_{i=1}\sum^3_{k=1}\left(\nabla_{\!k}v_i\,(\sigma^{ik}
+\tilde\sigma^{ik})+\nabla_{\!k}(\nabla_{\!i}T\,
\varkappa^{ik})\right).
\endgathered
\tag10.9
$$
Applying \thetag{9.4} to the equation \thetag{10.9}, we transform
it to the following one:
$$
\gathered
\rho\,T\,\frac{\partial s}{\partial t}
+\sum^3_{k=1}\rho\,T\,v^k\,\nabla_{\!k}s
+\sum^3_{i=1}\sum^3_{j=1}\sum^3_{k=1}
\frac{\bar\sigma^{ij}}{2}\!\left(\nabla_{\!i}v^k\,\hat G_{kj}+
\hat G_{ik}\,\nabla_{\!j}v^k\,-\right.\\
\left.-2\,\Theta_{ij}\right)
=\sum^3_{i=1}\sum^3_{k=1}\left(\nabla_{\!k}v_i\,(\sigma^{ik}
+\tilde\sigma^{ik})+\nabla_{\!k}(\nabla_{\!i}T\,
\varkappa^{ik})\right).
\endgathered
\quad
\tag10.10
$$
Now let's recall formula \thetag{6.20} and write analogous formula
in present case
$$
\hskip -2em
\sigma^{ik}=\sum^3_{j=1}\hat G^{\,i}_j\,\bar\sigma^{jk}.
\tag10.11
$$
Applying \thetag{10.11} to \thetag{10.10}, we simplify it substantially.
Here is resulting equation
$$
\gathered
T\,\frac{\partial(\,\rho\,s)}{\partial t}+\sum^3_{k=1}T\ 
\nabla_{\!k}(\,\rho\,s\,v^k)=\sum^3_{i=1}\sum^3_{j=1}
\sigma^{ij}\,\theta_{ij}\,+\\
+\sum^3_{i=1}\sum^3_{k=1}\left(\nabla_{\!k}v_i\,(\sigma^{ik}
+\tilde\sigma^{ik})+\nabla_{\!k}(\nabla_{\!i}T\,
\varkappa^{ik})\right).
\endgathered
\quad
\tag10.12
$$
For $\Theta_{ij}$ in \thetag{10.10} we used formula \thetag{9.8}.
By further transformations similar to that of section~7 we can
bring the equation \thetag{10.12} to the form analogous to
\thetag{7.11}:
$$
\gathered
\frac{\partial(\,\rho\,s)}{\partial t}+\sum^3_{k=1}\nabla_{\!k}\!
\left(\rho\,s\,v^k-\shave{\sum^3_{i=1}}\frac{\nabla_{\!i}T\,
\varkappa^{ik}}{T}\right)=\sum^3_{i=1}\sum^3_{j=1}\frac{\sigma^{ij}
\,\theta_{ij}}{T}\,+\\
+\sum^3_{i=1}\sum^3_{k=1}\sum^3_{j=1}\sum^3_{q=1}\frac{v_{ik}\,
\eta^{ikjq}\,v_{jq}}{T}+\sum^3_{i=1}\sum^3_{k=1}\frac{\nabla_{\!i}T
\,\varkappa^{ik}\,\nabla_{\!k}T}{T^2}.
\endgathered
\quad
\tag10.13
$$
As compared to \thetag{7.11} in \thetag{10.13} we have one extra term
in right hand side. It should be positive like other two terms in right
hand side of \thetag{10.13}:
$$
\hskip -2em
\sum^3_{i=1}\sum^3_{j=1}\frac{\sigma^{ij}\,\theta_{ij}}{T}\geqslant 0.
\tag10.14
$$
Formulas \thetag{10.13} and \thetag{10.14} yield physical interpretation
of tensor \thetag{9.9}. This tensor determines entropy production due
to plastic deformation of medium. \par
    Formula \thetag{10.11} relates tensor $\bar{\boldsymbol\sigma}$
and stress tensor $\boldsymbol\sigma$. Applying \thetag{10.11} to
\thetag{10.4}, we derive formula for stress tensor of isotropic
plastic medium:
$$
\hskip -2em
\gathered
\sigma^{ij}=f_{[1]}\ \hat G^{ij}+\sum^3_{k=1}\sum^3_{q=1}f_{[2]}\
\hat G^{ik}\,g_{kq}\,\hat G^{qj}\,+\\
+\,\sum^3_{k=1}\sum^3_{q=1}\sum^3_{m=1}\sum^3_{n=1}f_{[3]}\
\hat G^{ik}\,g_{kq}\,\hat G^{qm}\,g_{mn}\,\hat G^{nj}.
\endgathered
\tag10.15
$$
Formula \thetag{10.15} is analogous to \thetag{6.23}. Like formula
\thetag{6.23}, this formula means that stress tensor $\boldsymbol
\sigma$ is symmetric: $\sigma^{ij}=\sigma^{ji}$. In general case
for non-isotropic media symmetry of tensor $\boldsymbol\Pi$ in
\thetag{7.1} and hence symmetry of $\boldsymbol\sigma$ is derived
from conservation law for angular momentum (see \cite{4}).\par
    Note that in \thetag{9.2} we declared that tensors $\boldsymbol
\sigma$ and $\bar{\boldsymbol\sigma}$ depend only on elastic part
of deformation tensor \thetag{9.1}. However, for coefficients
$f_{[1]},\,f_{[2]},\,f_{[3]}$ in formulas \thetag{10.4} and
\thetag{10.15} we have explicit expression \thetag{10.5} containing
entry of density $\rho$. Due to \thetag{5.4} density of medium
depends on total deformation tensor $\bold G$, but not on its
elastic part $\hat{\bold G}$ only. In order to avoid this discrepancy
we need to introduce additional restriction for plastic part
of deformation tensor
$$
\hskip -2em
\det\check{\bold G}=1.
\tag10.16
$$
Then due to \thetag{9.1} and \thetag{10.16} we get $\det\hat{\bold G}
=\det\bold G$ and \thetag{5.4} is replaced by analogous relationship
binding density $\rho$ to elastic deformation tensor $\hat{\bold G}$:
$$
\hskip -2em
\ln\rho-\frac{\ln\det\hat{\bold G}-\ln\det\bold g}{2}=\const.
\tag10.17
$$
Restriction \thetag{10.16} and the equality \thetag{10.17} following
from it are reasonable from physical point of view. Indeed, according
to forgetting principle (see above), plastic deformations are those
which can be forgotten. However deformations changing density of medium
cannot be forgotten. Hence they cannot be purely plastic.\par
    Restriction \thetag{10.16} leads to the restriction for tensor
$\boldsymbol\theta$ in \thetag{9.12}. Indeed, differentiating
\thetag{10.16} and applying equation \thetag{9.12}, we derive
$$
\hskip -2em
\tr\boldsymbol\theta=\sum^3_{k=1}\theta^{\,k}_k=0.
\tag10.18
$$
Thus, formula \thetag{10.18} means that tensor $\boldsymbol\theta$
determines symmetric linear operator with zero trace. It is symmetric
due to \thetag{9.10} and symmetry $\theta_{kq}=\theta_{qk}$.\par
    Plastic deformation is a way for draining stress. Tensor
$\boldsymbol\theta$ determines the rate of such draining. In elastic
solid materials total deformation is purely elastic. Therefore
$\check G^{\,i}_j=\delta^i_j$. Substituting $\check G^{\,i}_j=\delta^i_j$
into the equations \thetag{9.12}, we find
$$
\hskip -2em
\theta^{\,i}_j=0.
\tag10.19
$$
Thus, elastic solid medium can be treated as limiting case of plastic
medium with vanishing tensor $\theta^{\,i}_j\to 0$.\par
    In liquids tensor $\boldsymbol\theta$ is undetermined. Indeed, above
in section~8 we treated liquid state media in the same way as purely
elastic solid state media, but with special form of free energy function
(see formula \thetag{8.2}). Then $\boldsymbol\theta$ is equal to zero
like in \thetag{10.19}. However, we could treat liquids as plastic media
by introducing some arbitrary tensor $\boldsymbol\theta$. In any case
stress tensor in liquids is determined by formula 
$$
\hskip -2em
\sigma^{ij}=-p\,g^{ij},
\tag10.20
$$
see \thetag{8.6} above. Substituting \thetag{10.20} into left hand
side of \thetag{10.14}, then taking into account \thetag{9.10} and
trace condition \thetag{10.18}, we derive
$$
\sum^3_{i=1}\sum^3_{j=1}\frac{\sigma^{ij}\,\theta_{ij}}{T}=
-\sum^3_{i=1}\sum^3_{j=1}\frac{p}{T}\,g^{ij}\,\theta_{ij}=
-\sum^3_{i=1}\frac{p}{T}\,\theta^{\,i}_i=0.
$$
This means that tensor $\boldsymbol\theta$ for liquid media is
not thermodynamically fixed. Therefore asymptotical behavior
of functions $\theta_{kq}=\theta_{kq}(T,\hat{\bold G})$ near phase
transition point from plastic solid state to liquid state
$T\to T_{\sssize\text{ph.tr.}}$ requires separate investigation.
This will be done in separate paper.

\enddocument
\end